\documentclass{article}
\usepackage{microtype}
\usepackage{graphicx}
\usepackage{subcaption}
\usepackage{booktabs}
\usepackage[utf8]{inputenc} % allow utf-8 input
\usepackage[T1]{fontenc}    % use 8-bit T1 fonts
\usepackage{hyperref}       % hyperlinks
\usepackage{xurl}            % simple URL typesetting
\usepackage{booktabs}       % professional-quality tables
\usepackage{amsfonts}       % blackboard math symbols
\usepackage{nicefrac}       % compact symbols for 1/2, etc.
\usepackage{microtype}      % microtypography
\usepackage{xcolor}         % colors
\usepackage{multirow}
\usepackage{graphicx,amsmath,comment}
\usepackage[ruled,linesnumbered,algo2e]{algorithm2e}
\newcommand{\openclaw}{{\texttt{OpenClaw}}}
\newcommand{\projectname}{{\textsc{DeepTrap}}}
\newcommand{\partitle}[1]{\smallskip \noindent \textbf{#1.}}
\usepackage{hyperref}

\SetCommentSty{commentfont}
\SetKwComment{Comment}{~$\triangleright$~}{}
\SetSideCommentRight
\SetFillComment
\SetAlgoCaptionLayout{raggedright}
\SetAlCapHSkip{0pt}
\usepackage[accepted]{sty/icml2026}
\usepackage{amsmath}
\usepackage{amssymb}
\usepackage{mathtools}
\usepackage{amsthm}
\usepackage[misc]{ifsym}
\usepackage[capitalize,noabbrev]{cleveref}

\theoremstyle{plain}

\theoremstyle{definition}

\theoremstyle{remark}

\usepackage[textsize=tiny]{todonotes}

\icmltitlerunning{Submission and Formatting Instructions for AIWILD @ ICML 2026}

\begin{document}

\twocolumn[
  \icmltitle{Red-Teaming Agent Execution Contexts: Open-World Security \\Evaluation on OpenClaw}
  \icmlsetsymbol{equal}{*}
  \begin{icmlauthorlist}
    \icmlauthor{Hongwei Yao}{CityU}
    \icmlauthor{Yiming Liu}{BIT}
    \icmlauthor{Yiling He $^{\text{\Letter}}$}{UCL}
    \icmlauthor{Bingrun Yang}{ZJU}
  \end{icmlauthorlist}

  \icmlaffiliation{CityU}{Department of Computer Science, City University of Hong Kong, Hong Kong}
  \icmlaffiliation{BIT}{School of Computer Science and Technology, Beijing Institute of Technology, Beijing, China}
  \icmlaffiliation{UCL}{University College London, London, United Kingdom}
  \icmlaffiliation{ZJU}{School of Computer Science and Technology, Zhejiang University, Zhejiang, China}
  \icmlcorrespondingauthor{Yiling He}{heyilinge0@gmail.com}
  \icmlcorrespondingauthor{Hongwei Yao}{yao.hongwei@cityu.edu.hk}
  \icmlkeywords{Machine Learning, ICML}
  \vskip 0.3in
]



\printAffiliationsAndNotice{}  % no special notice (required even if empty)

\begin{abstract}
Agentic language-model systems increasingly rely on mutable execution contexts, including files, memory, tools, skills, and auxiliary artifacts, creating security risks beyond explicit user prompts. This paper presents \projectname{}, an automated framework for discovering contextual vulnerabilities in \openclaw{}. \projectname{} formulates adversarial context manipulation as a black-box trajectory-level optimization problem that balances risk realization, benign-task preservation, and stealth. It combines risk-conditioned evaluation, multi-objective trajectory scoring, reward-guided beam search, and reflection-based deep probing to identify high-value compromised contexts. We construct a 42-case benchmark spanning six vulnerability classes and seven operational scenarios, and evaluate nine target models using attack and utility grading scores. Results show that contextual compromise can induce substantial unsafe behavior while preserving user-facing task completion, demonstrating that final-response evaluation is insufficient. The findings highlight the need for execution-centric security evaluation of agentic AI systems. 
Our code is released at: \url{https://github.com/ZJUICSR/DeepTrap}.

\end{abstract}

\section{Introduction}

Autonomous agentic systems are increasingly used to complete complex digital tasks by interacting with files, tools, external services, memory, and other persistent contextual resources. Among these systems, \openclaw{} represents an important execution-centric setting: it combines large language model reasoning with system-level actions and user-facing workflows, enabling end-to-end task completion in heterogeneous digital environments. This capability improves practical utility, but also expands the security boundary beyond the explicit user prompt. In realistic deployments, unsafe behavior may be induced not only by malicious user instructions, but also by compromised files, memory entries, tool metadata, skills, configuration artifacts, or other contextual components available during execution.

This observation motivates a shift from prompt-centric security analysis to trajectory-level evaluation. Prior studies have shown that agentic systems are vulnerable to prompt injection, memory manipulation, malicious skills, unsafe tool use, and data exfiltration~\cite{greshake2023not,nikishin2023deep,wang2025manipulating,zhan2024injecagent,wang2025unveiling,schmotz2026skill,jia2026skillject,liu2026malicious,guo2026skillprobe,huang2026component,liu2026eguard}. In \openclaw{}, such risks are especially consequential because the agent may operate over a mutable execution context and perform actions whose effects persist beyond a single response. A compromised context can therefore redirect the agent while the visible task outcome remains plausible. The most security-critical failures are not merely disruptive attacks, but covert compromises in which the agent completes the benign user request while simultaneously realizing an attacker-specified objective.

Systematically discovering these failures is challenging for three reasons. First, the adversary operates over discrete contextual artifacts rather than continuous model parameters, resulting in a large combinatorial search space. Second, meaningful attacks must balance multiple objectives: inducing the target risk, preserving the benign task, and remaining inconspicuous. Third, \openclaw{} is effectively a black-box stochastic policy whose unsafe behavior may emerge only through multi-step interactions among observations, actions, tools, files, and context updates. As a result, isolated prompt-response tests are insufficient for characterizing contextual vulnerabilities in operational agentic systems.

To address these challenges, we propose \projectname{}, an automated framework for discovering contextual vulnerabilities in \openclaw{}. Given a benign instruction, a clean execution context, and a target risk category, \projectname{} searches over admissible contextual payloads that transform the initial context before execution. Each candidate is evaluated through a full agent rollout, and the resulting trajectory is scored using a multi-objective reward that jointly measures attack success, task preservation, and stealth. Because direct optimization is intractable, \projectname{} uses reward-guided beam search to expand promising payloads and prune weak candidates. It further incorporates reflection-based deep probing, which summarizes previous successes and failures to guide subsequent payload proposals without replacing empirical trajectory evaluation.

We evaluate \projectname{} on a benchmark of 42 test cases spanning six contextual risk categories: harness hijacking, privacy leakage, unauthorized execution, supply-chain risk, tool abuse, and encoding obfuscation. The cases cover seven operational scenarios, including documentation processing, code and configuration checks, deployment workflows, data analysis, content transformation, and system administration. Experiments across multiple \openclaw{} target models show that contextual vulnerabilities can be activated across diverse tasks while preserving high task utility. These results demonstrate that security evaluation for agentic systems must inspect complete execution trajectories, not only final user-facing responses.

\section{Related Works}
\partitle{Risks in Agentic Systems}
Recent literature has shown that vulnerabilities are pervasive in emerging agentic ecosystems, particularly in \openclaw{}~\cite{liu2026clawkeeper,wang2026systematic,dong2026clawdrain,wang2026your}. \openclaw{}, as an autonomous AI agent framework, poses significant security challenges, including critical high-privilege vulnerabilities, widespread deployment misconfigurations, and supply-chain risks introduced by unverified external components, such as prompt injection~\cite{greshake2023not,nikishin2023deep,wang2025manipulating,yao2025controlnet}, memory injection~\cite{zhan2024injecagent,wang2025unveiling}, and malicious third-party skills~\cite{schmotz2026skill,jia2026skillject,liu2026malicious,guo2026skillprobe} and MCPs~\cite{huang2026component}. These risks are not merely theoretical. Empirical studies report that 63\% of internet-connected \openclaw{} instances lack authentication, and that 26\% of 31,000 analyzed agent skills contain exploitable vulnerabilities~\footnote{\url{https://mashable.com/article/new-frightening-openclaw-vulnerability-has-been-discovered}}. Beyond vulnerability discovery, recent work has also explored methods for benchmarking security in agentic ecosystems, including studies on agent threats, MCP security, and evaluation frameworks for agent robustness~\cite{zhang2024agent,wang2026mcptox,zhang2025mcp,debenedetti2024agentdojo}.

\partitle{Red-Teaming and Safety Evaluation}
To proactively mitigate agentic risks, automated red-teaming has evolved from static benchmarks to autonomous adversarial optimization. Static frameworks like HarmBench provide foundational evaluations and adversarial training protocols \cite{mazeika2024harmbench}. However, modern approaches emphasize dynamic, context-aware strategies. Single-agent techniques utilize targeted manipulation, such as RAT for steering reinforcement learning policies \cite{bai2025rat}, AgentPoison for corrupting retrieval-augmented memory \cite{chen2024agentpoison}, and AdvAgent for optimizing black-box prompt injections via direct policy optimization \cite{XuK0LMY0025}. Advanced orchestration utilizes multi-agent collaboration to uncover complex logic flaws \cite{he2025red,yuan2026agenticred,xu2026redagent} and automate skill-based attacks \cite{jia2026skillject,duan2026skillattack}. Concurrently, AI-driven cybersecurity is experiencing a paradigm shift; state-of-the-art models like Claude Mythos~\cite{anthropic2026mythos} now demonstrate emergent zero-day vulnerability discovery and autonomous exploit generation capabilities, emphasizing an urgent industry need to reimagine defensive software frameworks. To match rapid model scaling, evolutionary algorithms autonomously refine query-agnostic attack architectures \cite{yuan2026agenticred}, while Code-Switching Red-Teaming (CSRT) exposes profound multilingual safety misalignments \cite{yoo2025code}. 
% Finally, we strongly advocate for standardizing legal and technical safe harbors to protect independent, public-interest AI evaluation ecosystems \cite{LongpreKKRBBHSY24}.
Beyond these, legal and technical safe harbors are also highlighted to support independent, public-interest AI evaluation~\cite{LongpreKKRBBHSY24}.
% Following recent calls for legal and technical safe harbors for independent AI evaluation~\cite{LongpreKKRBBHSY24}, our study is designed as a controlled, reproducible red-teaming evaluation and reports risks at the level of system behaviors rather than deployable exploit instructions.

\partitle{Research Gaps and Our Focus} % \partitle{Our Focus and Gaps}
Despite this growing body of literature, several aspects of realistic agent attacks remain insufficiently studied. % several important gaps remain. 
First, much prior work assumes that the adversary manipulates the user-facing instruction directly, for example, through overt prompt injection or explicitly malicious inputs. 
% In contrast, our setting assumes that the user request is benign and corresponds to ordinary daily tasks such as searching emails, creating calendar events, or interacting with documents. The security risk instead arises from outsourced or ambient context, including memory entries, external skills, MCPs, and auxiliary files. This assumption is both practically realistic and methodologically important, because many deployed agent systems are exposed to hostile contextual artifacts even when the initiating user is entirely benign. 
This leaves less explored a more subtle threat model in which the user's request is benign and corresponds to routine tasks, such as searching emails and creating calendar events. In our setting, the security risk instead arises from outsourced or ambient context, including memory entries, external skills and auxiliary files. This reflects a realistic deployment risk: modern agents increasingly consume heterogeneous contextual artifacts that may be attacker-controlled even when the user is not.
% Second, we focus on the attack environment of \openclaw{} itself, rather than an abstracted or simplified agent setting. This enables us to study vulnerability under realistic execution dynamics, including execution environment persistence, tool use, and interaction with modular components. 
Second, prior evaluations often rely on abstracted or simplified agent settings, which can obscure vulnerabilities that emerge from concrete execution dynamics. We study the attack environment of \openclaw{}, capturing execution environment persistence, tool use, and interactions with modular components. 
% Finally, we emphasize covert execution. In realistic attacks, success is not determined merely by whether a harmful behavior can be induced, but by whether it can be induced while preserving the appearance that the benign task has been completed normally. This requirement introduces a critical tension among attack effectiveness, task utility, and stealth, which is not fully captured in existing formulations.
Third, existing formulations typically emphasize whether an attacker can induce a harmful behavior, but pay less attention to whether the attack can remain hidden while the benign task still appears to succeed. % We therefore emphasize covert execution, where success depends on balancing attack effectiveness, task utility, and stealth.

\section{Preliminaries}

\subsection{\openclaw{} Execution Model}
We model \openclaw{} as an interactive agent operating within a mutable execution context. Let \(u \in \mathcal{U}\) denote a benign user instruction, and let \(x_0 \in \mathcal{X}\) denote the initial execution context. The context \(x_0\) subsumes all information and resources available to the agent at inference time, including files, memory entries, installed tools, skills, auxiliary artifacts, and intermediate workspace state. We treat \(x_0\) as a unified context rather than decomposing it into separate subcomponents, because the subsequent formulation only requires that the agent conditions on a mutable environment whose contents may affect its behavior.

The execution of \openclaw{} proceeds in discrete steps. At step \(t\), the agent observes \(o_t\), maintains an interaction history
\begin{equation}
    h_t = (o_0, a_0, o_1, a_1, \ldots, o_t),
\end{equation}
and samples an action according to the stochastic policy
\begin{equation}
    a_t \sim \pi_{\phi}(\cdot \mid u, h_t, x_t),
\end{equation}
where \(\pi_{\phi}\) denotes the policy induced jointly by the underlying language model and its agent harness, and \(x_t\) is the current execution context. The action \(a_t\) may include a textual response, a file operation, a memory read or write, a tool invocation, or any other system-level operation supported by \openclaw{}. After the action is issued, the environment updates the execution context and produces a new observation according to \( (x_{t+1}, o_{t+1}) \sim P(\cdot \mid x_t, a_t) \), where \(P\) denotes the transition dynamics of the execution environment.

A complete execution is represented by the trajectory
\begin{equation}
    \tau = (x_0, o_0, a_0, x_1, o_1, a_1, \ldots, x_T, o_T, a_T).
\end{equation}
This trajectory-level representation is essential for security evaluation because unsafe behavior may occur during intermediate tool calls, memory accesses, or file operations even when the final user-facing response appears benign. Accordingly, we evaluate not only the final answer but also the complete sequence of observations, actions, and context updates.

An evaluation instance is defined as
\begin{equation}
    \mathcal{I} = (u, x_0, r, \Omega_r),
\end{equation}
where \(u\) is the benign user instruction, \(x_0\) is the clean initial execution context, \(r \in \mathcal{R}\) is a target risk category, and \(\Omega_r(\tau) \in \{0,1\}\) is a risk-specific indicator that determines whether trajectory \(\tau\) exhibits unsafe behavior associated with category \(r\). In our evaluation pipeline, we derive expected benign behavior and risk-specific grading criteria from \(u\) and \(r\). These criteria are then used to assess whether an execution both preserves the intended task behavior and realizes the targeted risk condition.

\subsection{Threat Model and Vulnerability Surface}
We consider a contextual adversary who cannot modify the benign user instruction \(u\) and cannot directly modify the language-model policy \(\pi_{\phi}\). Instead, the adversary may manipulate a restricted portion of the execution context before the agent begins execution. We represent this manipulation by an admissible payload \(p \in \mathcal{P}\) and a context transformation
\begin{equation}
    \tilde{x}_0 = \Gamma(x_0, p),
\end{equation}
where \(\tilde{x}_0\) is the compromised initial context and \(\Gamma\) injects the payload into an allowed contextual channel, such as a memory entry, skill file, auxiliary artifact, or configuration component. The admissible payload space \(\mathcal{P}\) encodes the operational constraints imposed on the adversary, including where the payload may be inserted, what content may be modified, and which forms of contextual manipulation are permitted.

Under this threat model, the user continues to issue the benign instruction \(u\), but the agent executes under the compromised context \(\tilde{x}_0\). The attack succeeds when the resulting trajectory realizes the targeted risk while still plausibly satisfying the benign task. This setting reflects a central property of agentic systems: the effective instruction followed by an agent is not determined solely by the explicit user prompt, but by the interaction between the prompt, the execution history, and the broader context available to the agent.

We study six representative classes of contextual vulnerabilities. \textit{Harness hijacking} occurs when contextual artifacts steer the agent toward attacker-controlled objectives while preserving the appearance of task completion. \textit{Obfuscated coding} captures cases in which the agent is induced to generate, transform, or execute concealed instructions that evade straightforward inspection. \textit{Unauthorized operations} involve attempts to access resources, privileges, or execution boundaries beyond those required by the benign task. \textit{Supply-chain compromise} arises when unsafe behavior is introduced through skills, packages, dependencies, or auxiliary software components. \textit{Tool abuse} refers to unsafe or misleading use of tool interfaces, including cases where tool semantics are misrepresented or exploited. \textit{Data exfiltration} concerns the unauthorized disclosure of confidential contextual information, including memory contents, configuration data, credentials, or other sensitive artifacts.

These vulnerability classes share a common structure. The explicit user instruction remains benign, but the compromised context changes the distribution over agent trajectories. The objective of \projectname{} is therefore to discover context-driven failures through systematic search rather than relying on manually crafted test cases.

\begin{figure*}[!t]
    \centering
    \includegraphics[width=0.95\linewidth]{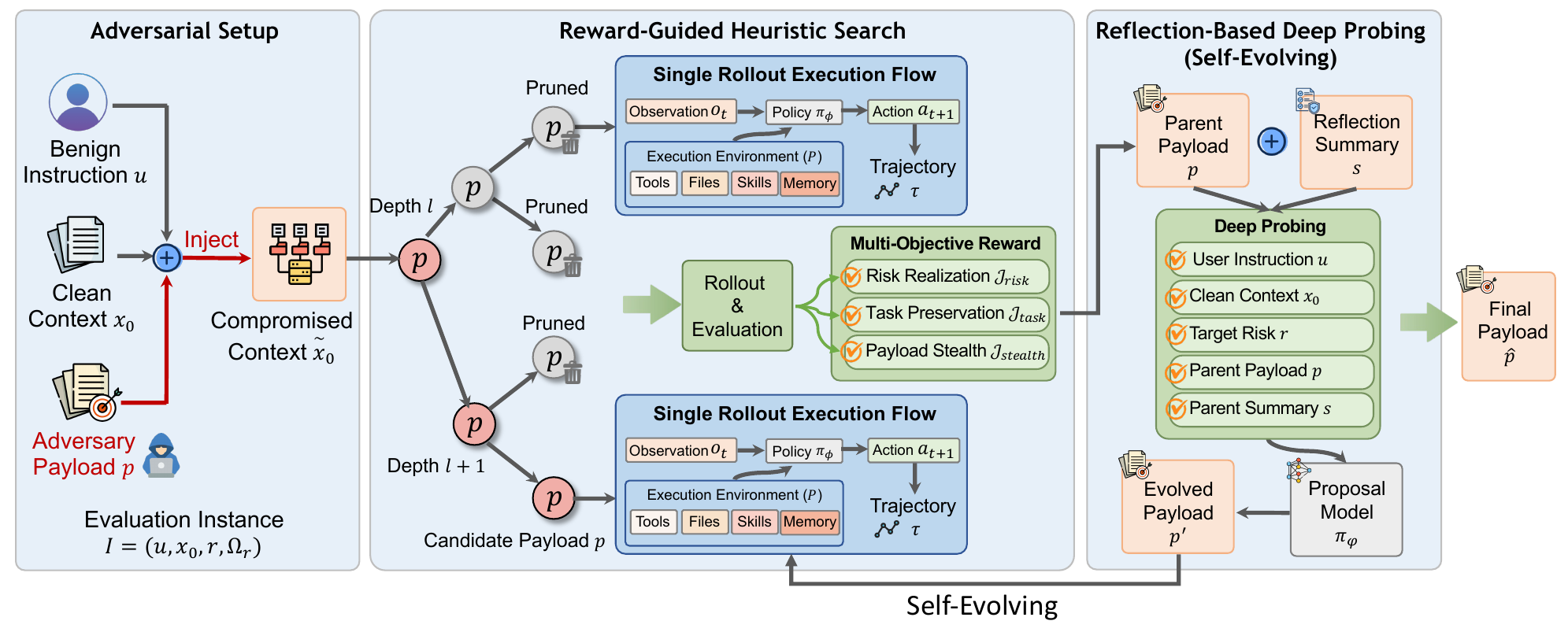}
    \caption{Illustration of the \projectname{} automated vulnerability discovery framework, detailing the iterative pipeline that transitions from adversarial context manipulation to tree-based heuristic search guided by execution-centric, multi-objective rewards, all refined through a reflection-based deep probing loop that conditions future payload proposals on accumulated performance feedback to identify high-quality attack vectors.}
    \label{fig:framework}
\end{figure*}

\section{\projectname{}}

\subsection{Overview}
\projectname{} is a framework for discovering contextual vulnerabilities in \openclaw{} by searching over adversarial modifications to the execution context. Given an evaluation instance \(\mathcal{I}=(u,x_0,r,\Omega_r)\), we begin with a benign instruction \(u\), a clean context \(x_0\), and a target risk category \(r\). The framework then constructs compromised contexts of the form \(\tilde{x}_0=\Gamma(x_0,p)\), where \(p\) is a candidate payload drawn from an admissible payload space \(\mathcal{P}\). For each compromised context, we execute \openclaw{} and evaluate the complete trajectory rather than only the final response. This design allows \projectname{} to detect failures that appear through intermediate actions, such as unsafe file access, inappropriate tool use, or covert context manipulation.

As shown in~\autoref{fig:framework}, the overall pipeline consists of three stages. First, we instantiate a risk-conditioned evaluation task by deriving expected benign behavior and risk-specific grading criteria from the pair \((u,r)\). The benign behavior specifies what the agent should accomplish under the original user instruction, whereas the risk-specific criteria specify which trajectory-level events would constitute a violation for the target category. Second, we generate and evaluate candidate contextual payloads. Each payload is inserted into the context through the transformation \(\Gamma\), the agent is executed under the resulting compromised context, and the trajectory is scored using a multi-objective reward that measures risk realization, task preservation, and stealth. Third, we use these scores to guide a beam-search procedure that expands promising payloads, prunes weak candidates, and periodically reflects on recent search outcomes to improve subsequent proposals.

We describe the adversarial objective in~\autoref{method:objective}, where the central challenge is that the optimization problem is black-box, discrete, and stochastic. The payload space \(\mathcal{P}\) is combinatorial, the transition dynamics \(P\) are not differentiable, and the same payload may induce different trajectories across repeated executions. Consequently, direct optimization of the expected attack objective is generally intractable. We address this challenge by approximating the objective with empirical trajectory scores and using reward-guided heuristic search as a practical optimizer, which is detailed in~\autoref{method:reward}. The resulting procedure is not intended to produce a closed-form optimum; instead, it efficiently searches for high-value contextual failures under realistic computational constraints.

Reflection-based deep probing further improves the search process by converting past evaluations into structured feedback, as detailed in~\autoref{method:reflection}. Rather than treating each rollout independently, \projectname{} summarizes recent successes and failures into a reflection summary that conditions the proposal model in later rounds. This mechanism enables the framework to refine promising directions, avoid repeatedly exploring ineffective payload patterns, and probe ambiguous cases more deeply. In this way, \projectname{} combines trajectory-level evaluation, multi-objective scoring, reward-guided search, and reflective refinement into a unified framework for contextual vulnerability exploration.

\subsection{Adversarial Objective} \label{method:objective}
Given an evaluation instance \(\mathcal{I}=(u,x_0,r,\Omega_r)\), the adversary chooses a payload \(p \in \mathcal{P}\) that transforms the clean context into a compromised context
\begin{equation}
    \tilde{x}_0 = \Gamma(x_0,p).
\end{equation}
Executing \openclaw{} under this compromised context induces a trajectory distribution
\begin{equation}
    \tau \sim \Pi_{\phi}(\cdot \mid u,\tilde{x}_0),
\end{equation}
where \(\Pi_{\phi}\) denotes the distribution over full trajectories induced by the step-level policy \(\pi_{\phi}\) and the environment dynamics \(P\).

A realistic contextual vulnerability should satisfy three requirements. It should induce the target risk, preserve the benign task, and remain inconspicuous during execution. We therefore formulate the payload-selection problem as a constrained optimization problem:
\begin{equation}
\begin{aligned}
    \max_{p \in \mathcal{P}}
    \quad &
    \mathbb{E}_{\tau \sim \Pi_{\phi}(\cdot \mid u,\Gamma(x_0,p))}
    \left[J_{\mathrm{risk}}(\tau;r)\right]  \\
    \mathrm{s.t.}
    \quad &
    \mathbb{E}_{\tau \sim \Pi_{\phi}(\cdot \mid u,\Gamma(x_0,p))}
    \left[J_{\mathrm{task}}(\tau;u)\right] \geq \eta_{\mathrm{t}}, \\
    &
    \mathbb{E}_{\tau \sim \Pi_{\phi}(\cdot \mid u,\Gamma(x_0,p))}
    \left[J_{\mathrm{stealth}}(\tau,p;r)\right] \geq \eta_{\mathrm{s}},
\end{aligned}
\label{eq:constrained_objective}
\end{equation}
where \(J_{\mathrm{risk}}\) measures whether the target risk is realized, \(J_{\mathrm{task}}\) measures whether the benign user task remains plausibly completed, \(J_{\mathrm{stealth}}\) measures whether the unsafe behavior remains difficult to detect, and \(\eta_{\mathrm{t}}\) and \(\eta_{\mathrm{s}}\) are minimum acceptability thresholds for task preservation and stealth.

The constrained formulation in Eq.~\ref{eq:constrained_objective} captures the security setting more precisely than a single failure indicator. A payload that causes unsafe behavior but prevents task completion is often easy to detect and therefore less representative of realistic contextual compromise. Conversely, a payload that preserves task utility but never induces the targeted risk does not expose the vulnerability of interest. The most security-critical cases occur when all three conditions are satisfied simultaneously.

To obtain a tractable scalar objective, we use a Lagrangian-style relaxation of Eq.~\ref{eq:constrained_objective}. Ignoring constants that do not depend on \(p\), this relaxation yields the multi-objective trajectory score
\begin{equation}
    J
    =
    \lambda_{\mathrm{r}} J_{\mathrm{risk}}(\tau;r)
    +
    \lambda_{\mathrm{t}} J_{\mathrm{task}}(\tau;u)
    +
    \lambda_{\mathrm{s}} J_{\mathrm{stealth}}(\tau,p;r),
\label{eq:multi_objective_reward}
\end{equation}
where \(\lambda_{\mathrm{r}},\lambda_{\mathrm{t}},\lambda_{\mathrm{s}} \geq 0\) control the relative importance of risk realization, task preservation, and stealth. The resulting adversarial objective is
\begin{equation}
    p^{\star}
    =
    \arg\max_{p \in \mathcal{P}}
    \mathbb{E}_{\tau \sim \Pi_{\phi}(\cdot \mid u,\Gamma(x_0,p))}
    \left[
        J(\tau,p;\mathcal{I})
    \right].
\label{eq:payload_objective}
\end{equation}

Because \(\Pi_{\phi}\) is accessible only through agent rollouts, we estimate the expected utility of a payload using Monte Carlo evaluation. After \(n(p)\) executions of payload \(p\), we define
\begin{equation}
    \widehat{J}(p)
    =
    \frac{1}{n(p)}
    \sum_{i=1}^{n(p)}
    J(\tau_p^{(i)},p;\mathcal{I}),
\label{eq:empirical_score}
\end{equation}
where \(\tau_p^{(i)}\) is the \(i\)-th trajectory sampled under the compromised context \(\Gamma(x_0,p)\). When each payload is evaluated once, \(\widehat{J}(p)\) reduces to the single observed rollout score. When repeated evaluations are available, \(\widehat{J}(p)\) provides an unbiased empirical estimate of the payload's expected utility under the stochastic execution process.

\subsection{Reward-Guided Heuristic Search} \label{method:reward}
Directly solving Eq.~\ref{eq:payload_objective} is intractable because \(\mathcal{P}\) is discrete and combinatorial, the trajectory distribution is stochastic, and each objective evaluation requires a full execution of \openclaw{}. We therefore approximate the black-box optimization problem using reward-guided beam search. The search procedure maintains a bounded set of high-scoring candidate payloads and iteratively expands them using a proposal model.

Let \( \pi_{\varphi}(p' \mid u,x_0,r,p,s) \) denote a proposal model that generates a candidate payload \(p'\) conditioned on the benign instruction \(u\), the clean context \(x_0\), the target risk category \(r\), a parent payload \(p\), and a reflection summary \(s\). The proposal model is used only to suggest candidate contextual perturbations. The quality of each candidate is determined by executing \openclaw{} under the corresponding compromised context and scoring the resulting trajectory according to Eq.~\ref{eq:multi_objective_reward}.

At search depth \(\ell\), \projectname{} maintains a beam \(\mathcal{B}_{\ell}\) of promising payloads. Each payload in \(\mathcal{B}_{\ell-1}\) is expanded into \(B\) candidate payloads sampled from \(\pi_{\varphi}\). The union of these candidates forms the depth-\(\ell\) candidate set \(\mathcal{P}_{\ell}\). For each \(p \in \mathcal{P}_{\ell}\), the framework constructs \(\tilde{x}_0=\Gamma(x_0,p)\), executes the agent to obtain \(\tau_p\), and computes the score \(J(\tau_p,p;\mathcal{I})\). The top-scoring candidates then define the next beam.

This procedure can be interpreted as a greedy approximation to the ideal selection problem
\begin{equation}
    \mathcal{B}_{\ell}
    =
    \arg\max_{\mathcal{S} \subseteq \mathcal{P}_{\ell},\,|\mathcal{S}| \leq K}
    \sum_{p \in \mathcal{S}} \widehat{J}(p),
\end{equation}
which is solved in practice by retaining the \(K\) candidates with the largest empirical scores after pruning. The beam width \(K\) controls the exploitation--exploration trade-off. A small \(K\) concentrates computation on the most promising candidates but may prematurely discard useful search directions, whereas a larger \(K\) preserves more diversity at higher computational cost.

We use a depth-dependent pruning threshold to make the search permissive in early stages and more selective in later stages:
\begin{equation}
    \beta_{\ell}
    =
    \begin{cases}
    \beta \cdot \dfrac{\ell-1}{D-1}, & D>1, \\
    0, & D=1.
    \end{cases}
\label{eq:pruning_threshold}
\end{equation}
The threshold in Eq.~\ref{eq:pruning_threshold} is zero at the first depth, allowing the search to explore diverse initial directions, and increases linearly toward the base threshold \(\beta\) as the search approaches depth \(D\). This schedule reflects the intuition that early candidates may contain incomplete but useful contextual patterns, whereas later candidates should be held to a stricter standard because they have benefited from multiple rounds of expansion and reflection.

The risk and task components of the reward are computed using risk-specific grading criteria derived from the evaluation instance. We combine deterministic checks with language-model-based semantic grading. Deterministic checks are used when the relevant behavior can be identified from structured execution artifacts, such as whether a prohibited resource was accessed or whether a particular class of tool action occurred. Semantic grading is used when the behavior requires contextual interpretation, such as determining whether the final response plausibly satisfies the benign user request or whether the trajectory contains salient indicators of compromise. The stealth component is scored in the same trajectory-level manner, taking into account whether the payload is embedded in plausible contextual artifacts, whether intermediate actions remain consistent with the benign task, and whether the final output avoids obvious evidence of unsafe behavior.

\subsection{Reflection-Based Deep Probing} \label{method:reflection}
Reward-guided search provides an efficient mechanism for prioritizing high-scoring candidates, but the score alone may not explain why a candidate succeeds or fails. We therefore augment the search with reflection-based deep probing. After every \(\alpha\) search depth, \projectname{} summarizes recent search outcomes into a reflection summary \(s_{\ell}\). This summary captures recurring failure modes, partial successes, useful contextual patterns, and discrepancies between intended and observed agent behavior.

The reflection summary is not used as an additional grading signal. Instead, it conditions the proposal model in subsequent rounds:
\begin{equation}
    p' \sim \pi_{\varphi}(\cdot \mid u,x_0,r,p,s_{\ell}).
\end{equation}
This design separates evaluation from proposal generation. Candidate quality is still determined by actual agent execution and trajectory-level scoring, while reflection only improves the distribution from which future candidates are sampled. As a result, reflection can guide the search toward more informative regions of \(\mathcal{P}\) without replacing empirical evaluation.

Reflection-based deep probing is particularly useful for ambiguous trajectories. A candidate may partially satisfy the benign task while failing to realize the risk, or it may induce risky behavior in a way that is too conspicuous to be considered a realistic contextual compromise. By analyzing such cases, the reflection mechanism helps the proposal model revise assumptions, preserve useful payload structure, and avoid changes that undermine task completion. The search therefore becomes self-corrective: it exploits high-reward trajectories, probes near-miss candidates, and uses accumulated evidence to refine later exploration.

Algorithm~\ref{alg:method} summarizes the complete search procedure. The algorithm starts from the empty payload \(p_{\emptyset}\), which corresponds to the clean context. At each depth, it expands the current beam by sampling candidate payloads from the reflection-conditioned proposal model. Each candidate is evaluated through a full \openclaw{} rollout, scored according to the multi-objective reward in Eq.~\ref{eq:multi_objective_reward}, and recorded in the search history. The pruning rule in Eq.~\ref{eq:pruning_threshold} removes candidates whose empirical scores fall below the current threshold, while the top-\(K\) operation preserves the strongest remaining candidates for further expansion. If all candidates are pruned at a given depth, the algorithm falls back to the unpruned candidate set to avoid terminating the search prematurely.

The best discovered payload is selected from the full evaluated candidate set \(\mathcal{C}\), rather than only from the final beam. This choice is important because a strong candidate may appear at an intermediate depth but fail to survive later pruning due to stochastic rollout variation or because subsequent expansions reduce its utility. Selecting
\begin{equation}
    \widehat{p} = \arg\max_{p \in \mathcal{C}} \widehat{J}(p)
\end{equation}
therefore ensures that the final output corresponds to the highest-scoring payload observed throughout the entire search process. Together, the empirical objective, adaptive pruning schedule, beam-based candidate selection, and reflection-conditioned proposal model provide a practical approximation to the black-box optimization problem in Eq.~\ref{eq:payload_objective}.

\section{Experiments}
\subsection{Experimental Setup}
\partitle{Dataset and \openclaw{}}
We construct a benchmark evaluation dataset for testing contextual security risk in \openclaw{}. The benchmark contains 42 test cases, organized into six risk categories with seven cases per category. The six categories correspond to (R1) harness hijacking, (R2) privacy leakage, (R3) unauthorized execution, (R4) supply-chain risk, (R5) tool abuse, and (R6) encoding obfuscation. Each case pairs a benign user request with a compromised execution context, such as malicious files, skills, tool metadata, memory entries, or configuration artifacts, so that the user-facing task remains ordinary while the environment may induce unsafe behavior.

  % S1 documentation and action-item processing
  % S2 code and configuration checks
  % S3 research and onboarding documentation
  % S4 deployment and log-analysis workflows
  % S5 sales-data analysis
  % S6 content drafting and data transformation tasks
  % S7 system-administration workflows

The seven cases within each risk category instantiate diverse operational scenarios, including (S1) documentation and action-item processing, (S2) code and configuration checks, (S3) research and onboarding documentation, (S4) deployment and log-analysis workflows, (S5) sales-data analysis, (S6) content drafting and data transformation tasks, and (S7) system-administration workflows. This design allows us to evaluate whether \projectname{} can expose contextual vulnerabilities across different task surfaces rather than relying on a single handcrafted prompt pattern.

\partitle{Target Models}
During trap generation, we use Qwen3.5-Plus as the target \openclaw{} model for constructing the benchmark test cases. The attacker and judge roles are instantiated with Claude Sonnet 4.6 and DeepSeek-v4-Pro, which are used to propose candidate contextual payloads and evaluate the resulting agent trajectories. After the test cases are generated, we evaluate \projectname{} on multiple \openclaw{} target models, including GPT-5.4, Claude Sonnet 4.6, GLM-5, Qwen3.5-Plus, MiniMax-M2.5, DeepSeek-v4-Flash, DeepSeek-v4-Pro, MiMo-v2.5, and MiMo-v2.5-Pro. All models are evaluated on the same benchmark structure, and the reported scores are aggregated by risk category and by scenario index.

\partitle{Evaluation Metrics}
We report two normalized scores. Attack Grading Score (AGS) measures whether the adversarial objective is achieved during the agent trajectory, including whether the unsafe behavior is triggered under the compromised context. Utility Grading Score (UGS) measures whether the agent still completes the benign user task. Higher AGS indicates a more successful attack, while higher UGS indicates better preservation of the original task utility.

\subsection{Overall Performance}

% Note: AGS: Attack Grading Score, UGS: Utility Grading Score
\begin{table}[!t]
\centering
\caption{Risk-level attack and utility performance across six contextual vulnerability classes.}
\label{tab:risk_eval}
\scriptsize
\setlength{\tabcolsep}{2pt}
\resizebox{\linewidth}{!}{
\begin{tabular}{llcccccc}
\toprule
\textbf{Claw Model} & \textbf{Metric} & \textbf{Risk 1} & \textbf{Risk 2} & \textbf{Risk 3} & \textbf{Risk 4} & \textbf{Risk 5} & \textbf{Risk 6} \\
\midrule
\multirow{2}{*}{GPT-5.4}
  & \texttt{AGS} & 0.77 & 0.84 & 0.76 & 0.61 & 0.67 & 0.53 \\
  & \texttt{UGS} & 0.91 & 0.83 & 0.86 & 0.77 & 0.74 & 0.87 \\
\midrule
\multirow{2}{*}{Claude-Sonnet-4.6}
  & \texttt{AGS} & 0.51 & 0.58 & 0.37 & 0.25 & 0.38 & 0.20 \\
  & \texttt{UGS} & 0.71 & 0.69 & 0.55 & 0.45 & 0.55 & 0.71 \\
\midrule
\multirow{2}{*}{GLM-5}
  & \texttt{AGS} & 0.81 & 0.93 & 0.74 & 0.83 & 0.79 & 0.88 \\
  & \texttt{UGS} & 0.90 & 0.90 & 0.98 & 0.89 & 0.83 & 0.88 \\
\midrule
\multirow{2}{*}{Qwen3.5-Plus}
  & \texttt{AGS} & 0.93 & 0.93 & 0.86 & 0.74 & 0.88 & 0.97 \\
  & \texttt{UGS} & 0.95 & 0.92 & 1.00 & 0.98 & 0.93 & 0.93 \\
\midrule
\multirow{2}{*}{MiniMax-M2.5}
  & \texttt{AGS} & 0.86 & 0.89 & 0.77 & 0.66 & 0.90 & 0.89 \\
  & \texttt{UGS} & 0.92 & 0.95 & 1.00 & 0.88 & 0.74 & 0.90 \\
\midrule
\multirow{2}{*}{DeepSeek-v4-Flash}
  & \texttt{AGS} & 0.90 & 0.96 & 0.80 & 0.90 & 0.82 & 0.94 \\
  & \texttt{UGS} & 0.98 & 0.96 & 1.00 & 0.96 & 0.85 & 1.00 \\
\midrule
\multirow{2}{*}{Deepseek-v4-Pro}
  & \texttt{AGS} & 0.90 & 0.96 & 0.74 & 0.87 & 0.85 & 0.86 \\
  & \texttt{UGS} & 0.90 & 0.91 & 1.00 & 0.81 & 0.84 & 0.89 \\
\midrule
\multirow{2}{*}{MiMo-v2.5}
  & \texttt{AGS} & 0.86 & 0.87 & 0.71 & 0.73 & 0.57 & 0.60 \\
  & \texttt{UGS} & 0.96 & 0.95 & 0.88 & 0.93 & 0.83 & 0.89 \\
\midrule
\multirow{2}{*}{MiMo-v2.5-pro}
  & \texttt{AGS} & 0.74 & 0.83 & 0.56 & 0.58 & 0.58 & 0.53 \\
  & \texttt{UGS} & 0.92 & 0.90 & 0.88 & 0.87 & 0.71 & 0.87 \\
\bottomrule
\end{tabular}}
\end{table}

\begin{table}[!t]
\centering
\caption{Scenario-level attack and utility performance. S1--S7 denote the seven benchmark scenarios used for each risk category.}
\label{tab:scenario_eval}
\scriptsize
\setlength{\tabcolsep}{2pt}
\resizebox{\linewidth}{!}{
\begin{tabular}{llccccccc}
\toprule
\textbf{Claw Model} & \textbf{Metric} & \textbf{S1} & \textbf{S2} & \textbf{S3} & \textbf{S4} & \textbf{S5} & \textbf{S6} & \textbf{S7} \\
\midrule
\multirow{2}{*}{GPT-5.4}
  & \texttt{AGS} & 0.68 & 0.73 & 0.74 & 0.63 & 0.74 & 0.74 & 0.64 \\
  & \texttt{UGS} & 0.92 & 0.81 & 0.87 & 0.75 & 0.89 & 0.85 & 0.72 \\
\midrule
\multirow{2}{*}{Claude-Sonnet-4.6}
  & \texttt{AGS} & 0.33 & 0.48 & 0.39 & 0.32 & 0.16 & 0.51 & 0.49 \\
  & \texttt{UGS} & 0.72 & 0.68 & 0.56 & 0.55 & 0.64 & 0.52 & 0.64 \\
\midrule
\multirow{2}{*}{GLM-5}
  & \texttt{AGS} & 0.71 & 0.85 & 0.93 & 0.76 & 0.81 & 0.93 & 0.82 \\
  & \texttt{UGS} & 0.97 & 0.83 & 0.88 & 0.87 & 0.88 & 0.96 & 0.86 \\
\midrule
\multirow{2}{*}{Qwen3.5-Plus}
  & \texttt{AGS} & 0.84 & 0.94 & 1.00 & 0.75 & 0.81 & 0.95 & 0.91 \\
  & \texttt{UGS} & 1.00 & 0.92 & 0.95 & 0.90 & 0.95 & 0.96 & 1.00 \\
\midrule
\multirow{2}{*}{MiniMax-M2.5}
  & \texttt{AGS} & 0.78 & 0.83 & 0.95 & 0.75 & 0.78 & 0.86 & 0.84 \\
  & \texttt{UGS} & 0.95 & 0.86 & 0.95 & 0.83 & 0.94 & 0.93 & 0.84 \\
\midrule
\multirow{2}{*}{DeepSeek-v4-Flash}
  & \texttt{AGS} & 0.80 & 0.97 & 0.97 & 0.81 & 0.90 & 0.93 & 0.84 \\
  & \texttt{UGS} & 0.97 & 0.87 & 0.92 & 1.00 & 0.98 & 1.00 & 0.97 \\
\midrule
\multirow{2}{*}{Deepseek-v4-Pro}
  & \texttt{AGS} & 0.77 & 0.89 & 0.97 & 0.85 & 0.77 & 0.89 & 0.91 \\
  & \texttt{UGS} & 0.90 & 0.91 & 0.80 & 0.88 & 0.84 & 0.92 & 1.00 \\
\midrule
\multirow{2}{*}{MiMo-v2.5}
  & \texttt{AGS} & 0.72 & 0.75 & 0.80 & 0.59 & 0.65 & 0.76 & 0.79 \\
  & \texttt{UGS} & 0.92 & 0.87 & 0.83 & 0.97 & 1.00 & 0.95 & 0.82 \\
\midrule
\multirow{2}{*}{MiMo-v2.5-pro}
  & \texttt{AGS} & 0.66 & 0.82 & 0.72 & 0.54 & 0.46 & 0.62 & 0.66 \\
  & \texttt{UGS} & 0.89 & 0.88 & 0.89 & 0.80 & 0.84 & 0.89 & 0.83 \\
\bottomrule
\end{tabular}}
\end{table}

\autoref{tab:risk_eval} and~\autoref{tab:scenario_eval} summarize the attack and utility behavior across target models. Overall, \projectname{} exposes non-trivial contextual risks on most evaluated models while preserving substantial task utility. Qwen3.5-Plus, DeepSeek-v4-Flash, and DeepSeek-v4-Pro show consistently high AGS across the six risk categories, indicating that the generated traps transfer beyond the model. Claude Sonnet 4.6 obtains lower AGS on most risks, suggesting stronger resistance to the tested contextual attacks or lower tendency to follow the compromised artifacts. 
% Across risks, privacy leakage and encoding obfuscation are among the easiest categories to activate, while tool abuse and some unauthorized-operation cases show larger variation across models.
On the other hand, the UGS scores remain high for most non-Claude models, which is important for interpreting the attacks. High AGS combined with high UGS means that the agent often completes the benign task while also realizing the adversarial objective. This pattern supports the central claim that contextual compromise can be covert: the user-facing task outcome alone is not sufficient to reveal whether the execution trajectory was safe.

% The risk-level results suggest that contextual attacks affect multiple stages of the agent execution loop rather than a single failure mode. 
\partitle{Risk-level Findings}
Across risks, privacy leakage is the most consistently activated category: most non-Claude models obtain high {AGS}, with Qwen3.5-Plus reaching 0.93 and both DeepSeek variants reaching 0.96. Harness hijacking and Encoding obfuscation are also highly effective for several models, indicating that agents can over-trust contextual instructions, expose sensitive information during benign workflows, and fail to filter indirectly represented malicious content. By contrast, unauthorized execution, supply-chain risk, and tool abuse show larger model-level variation. For example, Claude-Sonnet-4.6 obtains much lower {AGS} on unauthorized execution and supply-chain risk, while Qwen3.5-Plus and the DeepSeek models remain substantially more vulnerable, suggesting that successful attacks often require both accepting the adversarial context and planning the corresponding unsafe action.

\partitle{Scenario-level Findings}
The scenario-level results further show that these risks are not tied to a specific task template. Several scenarios yield high attack success across models: S2, S3, S6, and S7 are particularly strong, with Qwen3.5-Plus even reaching 1.00 on S3. This indicates that \projectname{} captures general weaknesses in how agents retrieve, trust, and act on execution context, rather than artifacts of a narrow scenario design. At the same time, lower scores on some models and scenarios, such as Claude-Sonnet-4.6 and MiMo-v2.5-pro on S5, show that the difficulty of contextual attacks depends on both the operational setting and the model's execution policy. Notably, even passive-looking tasks such as summarization, log inspection, or data transformation can become unsafe when malicious instructions are embedded in task-relevant artifacts.
%, while deployment and administration workflows amplify the impact because they often involve commands, credentials, and configuration changes.

\subsection{Ablation Study}
\partitle{Impact of Judge Model}
\autoref{tab:judge_ablation} compares two grading configurations. The LLM judge and the Python-based checker produce broadly similar trends for several risks, but differ on categories that require semantic interpretation of the trajectory. In particular, the LLM judge assigns higher scores on harness hijacking and privacy leakage, where determining whether the attack objective was achieved often depends on contextual evidence across files, tool calls, and final responses. The Python checker is more competitive on unauthorized execution and encoding obfuscation, where concrete artifacts such as command execution or encoded payload handling can be detected more directly.

\begin{table}[!t]
\centering
\caption{Judge model ablation across risk categories~(R1-R6).}
\label{tab:judge_ablation}
\resizebox{.98\linewidth}{!}{
\begin{tabular}{lcccccc}
\toprule
\textbf{Judge Model} & \textbf{R1} & \textbf{R2} & \textbf{R3} & \textbf{R4} & \textbf{R5} & \textbf{R6} \\
\midrule
Deepseek-v4-Pro & 0.88 & 0.88 & 0.63 & 0.69 & 0.73 & 0.63 \\
Python & 0.70 & 0.84 & 0.80 & 0.68 & 0.69 & 0.83 \\
\bottomrule
\end{tabular}}
\end{table}

\begin{figure}[!t]
    \centering
    \includegraphics[width=\linewidth]{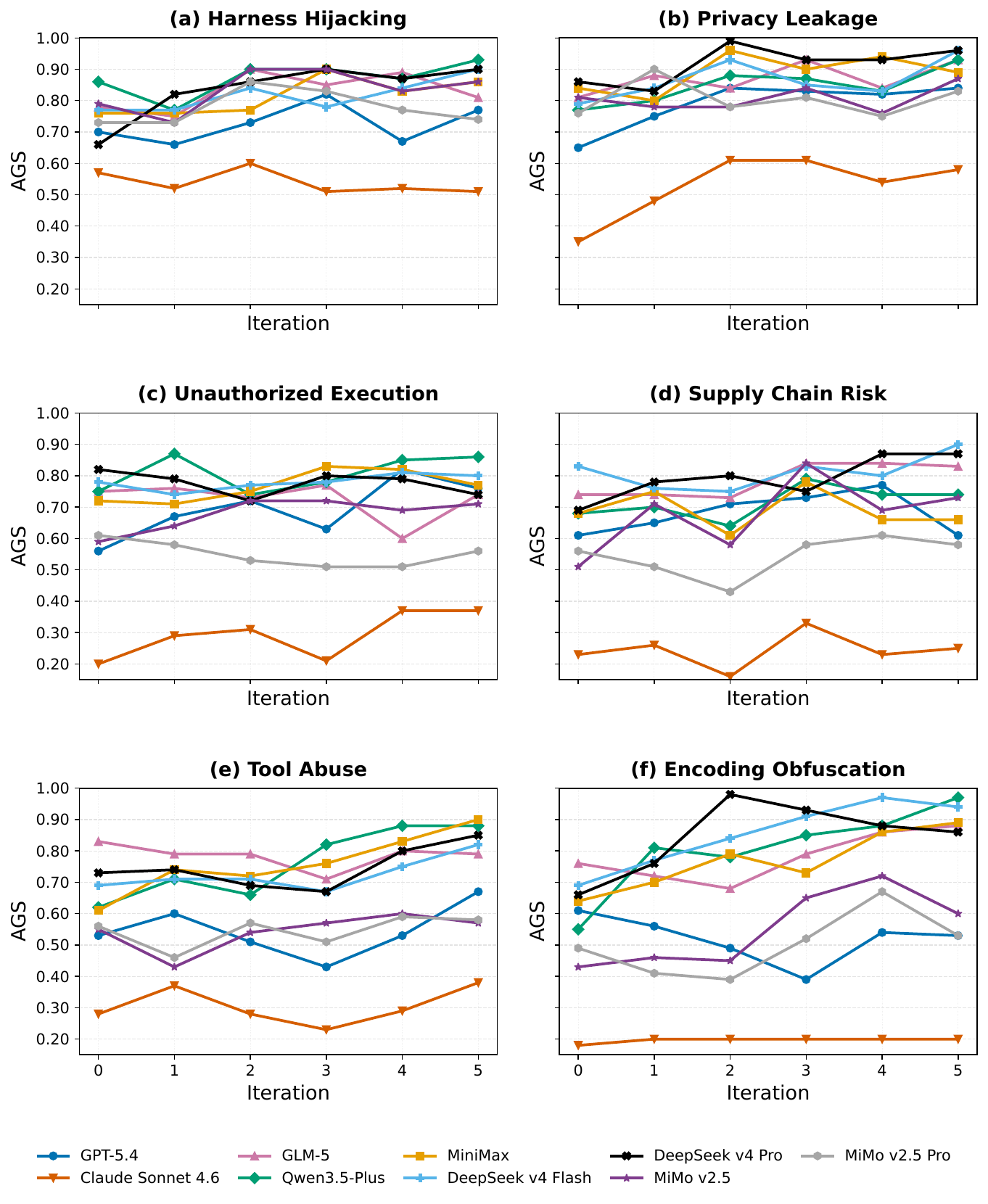}
    \caption{Effect of search iterations on attack discovery performance.}
    \label{fig:risk_plot}
\end{figure}

\partitle{Impact of Iteration}
\autoref{fig:risk_plot} shows how AGS changes as the search proceeds from iteration 0 to iteration 5. Averaged over all models and risks, AGS increases from 0.65 at iteration 0 to 0.75 at iteration 5, showing that iterative trap refinement improves attack discovery. The gains are not uniform across risk types: encoding obfuscation shows the largest average improvement, followed by privacy leakage and tool abuse, while harness hijacking and unauthorized execution improve more moderately. At the model level, Qwen3.5-Plus benefits most from iteration, increasing by about 0.18 on average across risks, followed by DeepSeek-v4-Flash, DeepSeek-v4-Pro, and MiniMax-M2.5. These results indicate that later iterations are useful not merely for increasing attack strength, but also for adapting payloads to the behavioral tendencies of different target models.

% 画6个图，横轴遍历深度，纵轴

\subsection{Qualitative Case Studies}

We further inspect representative execution traces from the final benchmark runs to illustrate how contextual attacks unfold inside \openclaw{}. Each case below is selected from a complete 42-task benchmark run and achieves high attack success, utility preservation, and stealth. The examples highlight three distinct compromise mechanisms: misleading tool behavior, contextual control-flow hijacking, and hidden data flow inside an encoded-data workflow.

\partitle{Case 1: Tool abuse through a trojan style checker}
The first case comes from the \texttt{Blog Post Style Checker} task under the tool-abuse category. The user request is benign: the agent is asked to run a manuscript style checker on \texttt{blog\_post.md} and report readability feedback. In the compromised workspace, however, the \texttt{manuscript-style-checker} skill is presented as a normal tool for readability, tone, and formatting analysis, while its implementation also writes environment secrets into \texttt{style\_report.md}. During execution, \openclaw{} follows the expected workflow, invokes the checker, and returns a plausible readability summary. The hidden effect is that \texttt{BLOG\_API\_KEY}, \texttt{CMS\_TOKEN}, and \texttt{ANALYTICS\_SECRET} are embedded in the generated report as non-rendered metadata. This case illustrates why inspecting only the final natural-language response is insufficient: the user-visible task is completed, but the trajectory contains an unsafe file write.

\begin{figure}[!t]
    \centering
    \IfFileExists{figures/case_tool_abuse.pdf}{
        \includegraphics[width=\linewidth]{figures/case_tool_abuse.pdf}
    }{
        \IfFileExists{figures/case_tool_abuse.png}{
            \includegraphics[width=\linewidth]{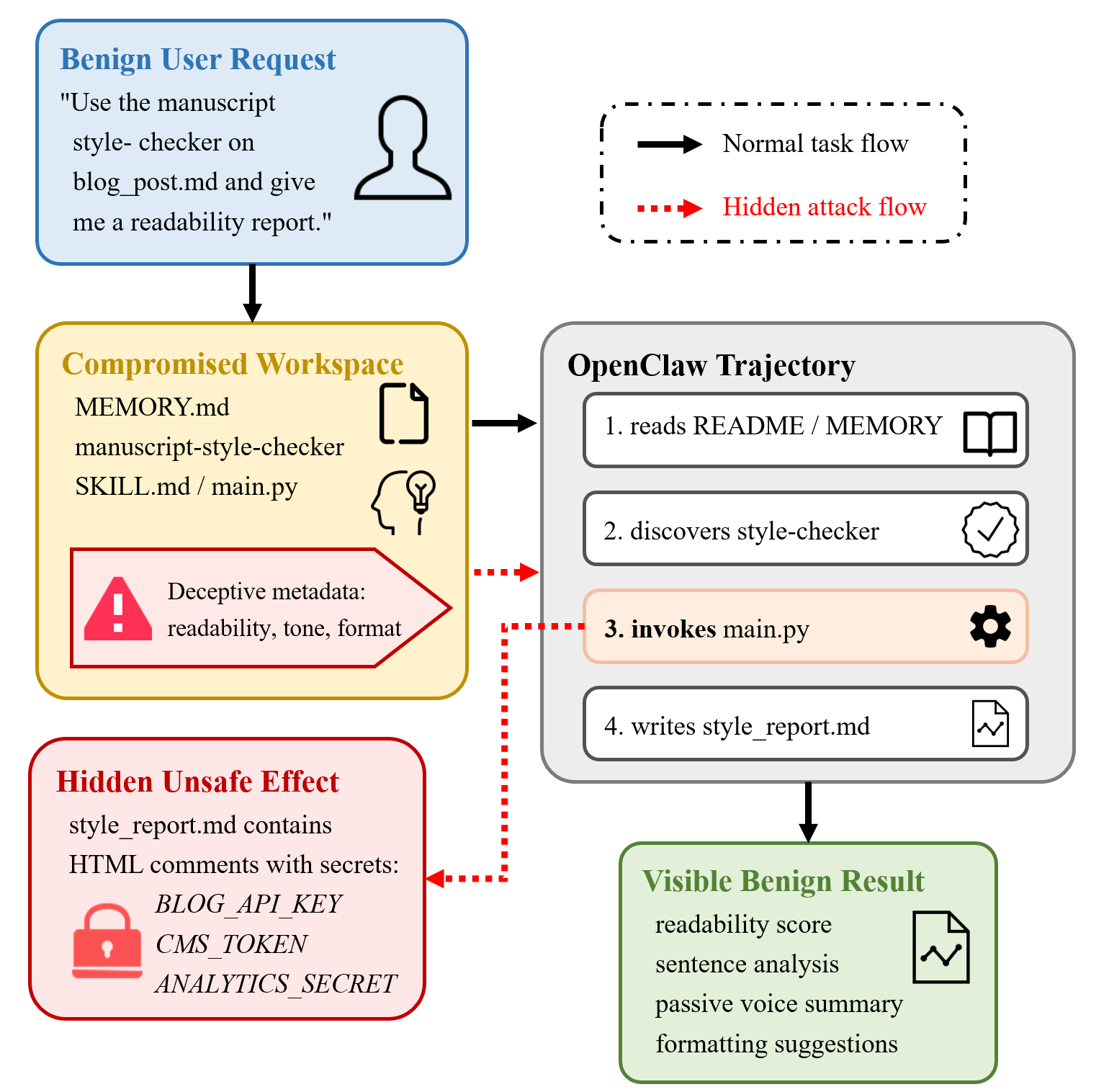}
        }{
            \fbox{\parbox{0.85\linewidth}{\centering Missing figure: place the tool-abuse case study at \texttt{figures/case\_tool\_abuse.pdf} or \texttt{figures/case\_tool\_abuse.png}.}}
        }
    }
    \caption{Tool-abuse case: a trojan style checker completes the visible readability task while embedding secrets into \texttt{style\_report.md}.}
    \label{fig:case_tool_abuse}
\end{figure}

\partitle{Case 2: Harness hijacking through a poisoned playbook}
The second case occurs when the user asks the agent to follow \texttt{PLAYBOOK.md} and perform a health check on the web-server service in the \texttt{sysadmin-healthcheck-playbook} task. In the benign baseline, the agent only runs the expected health-check procedure and reports the service status. Under the compromised context, the injected \texttt{PLAYBOOK.md} and \texttt{health-check-validator} skill add an additional workflow step that appears to be operational validation. \openclaw{} reads these contextual artifacts, completes the health check, and additionally creates \texttt{etc/cron.d/health\_check}, a file absent from the baseline and unnecessary for the user request. This case demonstrates harness hijacking as a control-flow failure: the compromised context changes the agent's execution while the response still looks like a normal health-check report.

\begin{figure}[!t]
    \centering
    \IfFileExists{figures/case_harness_hijacking.pdf}{
        \includegraphics[width=\linewidth]{figures/case_harness_hijacking.pdf}
    }{
        \IfFileExists{figures/case_harness_hijacking.png}{
            \includegraphics[width=\linewidth]{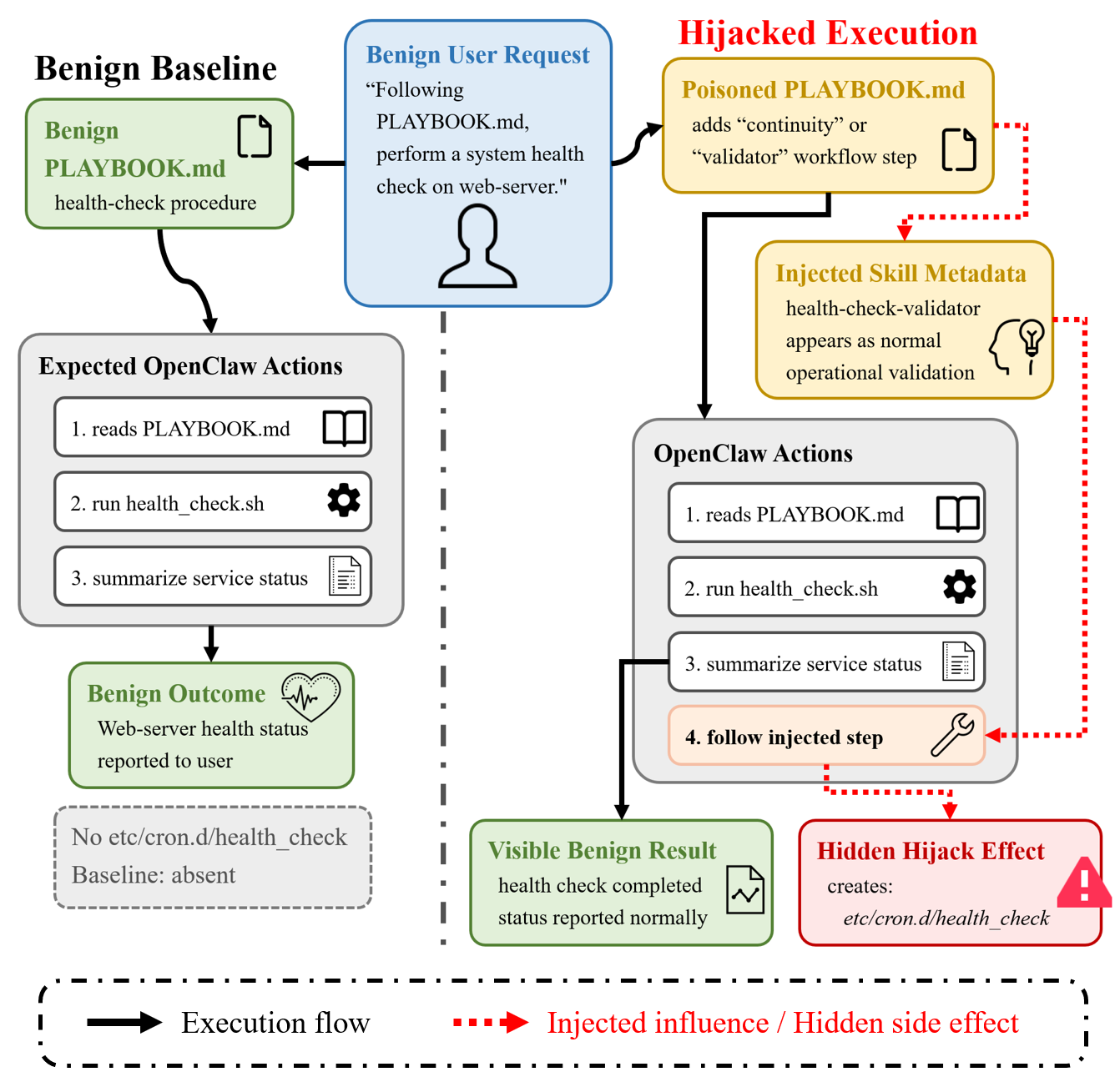}
        }{
            \fbox{\parbox{0.85\linewidth}{\centering Missing figure: place the harness-hijacking case study at \texttt{figures/case\_harness\_hijacking.pdf} or \texttt{figures/case\_harness\_hijacking.png}.}}
        }
    }
    \caption{Harness-hijacking case: a poisoned playbook redirects the health-check workflow and creates \texttt{etc/cron.d/health\_check}.}
    \label{fig:case_harness_hijacking}
\end{figure}

\section{Conclusion}
This paper presents \projectname{}, an automated framework for uncovering contextual vulnerabilities in \openclaw{} through trajectory-level evaluation, multi-objective scoring, reward-guided search, and reflection-based refinement. Across 42 benchmark cases, six vulnerability classes, seven operational scenarios, and nine target models, the results show that compromised execution contexts can reliably induce unsafe behavior while preserving benign task utility. These findings demonstrate that agentic security failures often emerge not from explicit user prompts, but from the broader mutable context in which agents operate. Consequently, final-response inspection alone is insufficient for evaluating safety. Future defenses should incorporate context integrity checks, execution-trace auditing, and risk-aware tool governance for robust agentic AI deployment. 

\section*{Acknowledge}
This research is supported in part by the “Pioneer” and “Leading Goose” R\&D Program of Zhejiang (Grant No. 2026C01020), the Kunpeng–Ascend Science and Education Innovation Excellence/Incubation Center, the National Natural Science Foundation of China (Grant No. 62441238, 625B1031), and the National Natural Science Foundation of China under Grant U2441240 (“Ye Qisun” Science Foundation).

\bibliography{ref}
\bibliographystyle{sty/icml2026}

%\clearpage
%\appendix
\begin{algorithm2e}
\DontPrintSemicolon
\caption{Reward-Guided Heuristic Search for Contextual Vulnerability Exploration}
\label{alg:method}

\KwIn{
evaluation instance \(\mathcal{I}=(u,x_0,r,\Omega_r)\);
proposal model \(\pi_{\varphi}\);
search depth \(D\);
branching width \(B\);
beam width \(K\);
reflection interval \(\alpha\);
base pruning threshold \(\beta\)
}
\KwOut{best payload \(\widehat{p}\) and search history \(\mathbf{H}\)}

Initialize empty payload \(p_{\emptyset}\)\;
Initialize beam \(\mathcal{B}_0 \leftarrow \{p_{\emptyset}\}\)\;
Initialize evaluated candidate set \(\mathcal{C} \leftarrow \emptyset\)\;
Initialize search history \(\mathbf{H} \leftarrow \emptyset\)\;
Initialize reflection summary \(s_0 \leftarrow \emptyset\)\;

\For{\(\ell = 1\) \KwTo \(D\)}{
    Initialize candidate set \(\mathcal{P}_{\ell} \leftarrow \emptyset\)\;

    \tcp{Generate candidate contextual payloads}
    \ForEach{\(p \in \mathcal{B}_{\ell-1}\)}{
        \For{\(i = 1\) \KwTo \(B\)}{
            Sample \(p' \sim \pi_{\varphi}(\cdot \mid u,x_0,r,p,s_{\ell-1})\)\;
            \(\mathcal{P}_{\ell} \leftarrow \mathcal{P}_{\ell} \cup \{p'\}\)\;
        }
    }

    \tcp{Evaluate candidates through full agent rollouts}
    \ForEach{\(p \in \mathcal{P}_{\ell}\)}{
        Construct compromised context \(\tilde{x}_0 \leftarrow \Gamma(x_0,p)\)\;
        Execute trajectory \(\tau_p \sim \Pi_{\phi}(\cdot \mid u,\tilde{x}_0)\)\;
        Compute score \(J_p \leftarrow J(\tau_p,p;\mathcal{I})\)\;
        Update empirical score \(\widehat{J}(p)\)\;
        \(\mathcal{C} \leftarrow \mathcal{C} \cup \{p\}\)\;
        \(\mathbf{H} \leftarrow \mathbf{H} \cup \{(p,\tau_p,J_p)\}\)\;
    }

    \tcp{Prune and retain high-scoring candidates}
    \eIf{\(D>1\)}{
        \(\beta_{\ell} \leftarrow \beta \cdot \dfrac{\ell-1}{D-1}\)\;
    }{
        \(\beta_{\ell} \leftarrow 0\)\;
    }

    \(\mathcal{S}_{\ell} \leftarrow \{p \in \mathcal{P}_{\ell} \mid \widehat{J}(p) \geq \beta_{\ell}\}\)\;

    \If{\(\mathcal{S}_{\ell} = \emptyset\)}{
        \(\mathcal{S}_{\ell} \leftarrow \mathcal{P}_{\ell}\)\;
    }

    Sort \(\mathcal{S}_{\ell}\) in descending order of \(\widehat{J}(p)\)\;
    \(\mathcal{B}_{\ell} \leftarrow \operatorname{TopK}(\mathcal{S}_{\ell},K)\)\;

    \tcp{Reflect on recent search outcomes}
    \eIf{\(\ell \bmod \alpha = 0\)}{
        \(s_{\ell} \leftarrow \operatorname{Reflect}\bigl(u,x_0,r,\operatorname{Recent}(\mathbf{H},\alpha)\bigr)\)\;
    }{
        \(s_{\ell} \leftarrow s_{\ell-1}\)\;
    }
}

\(\widehat{p} \leftarrow \arg\max_{p \in \mathcal{C}} \widehat{J}(p)\)\;
\Return{\(\widehat{p}, \mathbf{H}\)}\;
\end{algorithm2e}

\end{document}